# The impact of incoming preparation and demographics on performance in Physics I: a multi-institution comparison

Shima Salehi [a, b], Eric Burkholder [a], Peter Lepage [c], Steven Pollock [d], and Carl Wieman [a, b]

[a] Department of Physics, Stanford University, Stanford, CA, USA 94305
[b] Graduate School of Education, Stanford University, Stanford, CA, USA  94305
[c] Department of Physics, Cornell University, Ithaca, NY, USA 14853
[d] Department of Physics, University of Colorado Boulder, Boulder, CO, USA 80309

## Abstract

We have studied the impact of incoming preparation and demographic variables on student performance on the final exam in physics 1, the standard introductory calculus-based mechanics course. This was done at three different institutions using multivariable regression analysis to examine the extent to which exam scores can be predicted by a variety of variables that are available to most faculty and departments. We have found that the results are surprisingly consistent across the institutions, with the only two variables that have predictive power being math SAT/ACT scores and concept inventory pre-scores. The importance of both variables is comparable and fairly similar across the institutions. They explain 20-30% of the variation in students' performance on the final exam. Most notably, the demographic variables (gender, under-represented minority, first generation to attend college) are not significant. In all cases, although there appear to be gaps in exam performance if one considers only the demographic variable, once these two proxies of incoming preparation are included in the model, there is no longer a demographic gap. There is only a preparation gap that applies equally across the entire student population. This work shows that to properly understand differences in student performance across a diverse population, and hence to design more effective instruction, it is important to do statistical analyses that take multiple variables into account.  It also illustrates the importance of having measures that are sensitive to both subject specific and more general preparation. The results suggest that better matching of the course design and teaching to the incoming student preparation will likely be the most effective way to eliminate observed performance gaps across demographic groups while also improving the success of all students.

## Introduction

Physics education researchers have made great progress in finding teaching methods that result in improvements in student learning when looking at class averages [1-4]. A recent and growing focus has been to go beyond averages and overall normalized gains, to look at how teaching methods impact different students in different ways [5-9]. This is an important step in finding how to best serve the different student subpopulations in our classes, including providing inclusive learning environments for historically under-represented demographic populations in STEM fields. That is an essential step for improving the diversity in physics in particular, and STEM fields in general. The first step in such research is to identify which factors are important in determining student outcomes for different populations, and hence, where it would be most effective to focus teaching improvements and research.

We use data from three different institutions to explore the effect of a variety of student characteristics on their score on the final exam in the large introductory calculus-based physics course ("physics 1"). Nearly all prospective engineering students as well as many science students take this course, and students' academic performance in this course is consequential in pursuing STEM majors in their undergraduate studies. This work does not consider all factors that might be important, but rather a set that most physics instructors or departments will have access to, e.g. incoming SAT or ACT scores, demographic information (gender, first generation (FG) status, and under-represented minority (URM) status), and pre-course physics concept inventory (CI) scores.

Of particular concern to many institutions today are the average gaps in performance often seen between different demographic groups, such as course grades, exam scores, and passing rates [10-17]. Underperformance of demographically under-represented students in physics 1 can have considerable negative influence on their prospect of pursuing STEM fields, thereby preventing the increase of their representation in those fields. The factors that give rise to such gaps and how we can best design learning environments to address them are important unanswered questions. It is important to identify and remove factors that might produce such gaps, but there is also a danger associated with focusing on such gaps. There are negative consequences to labeling gaps as demographic gaps when the gaps are not arising from demographic status per se, but from the factors correlated with it. This mislabeling can result in bias and negative expectations for the labeled demographic group by both instructors and students [18-20].

The most important result of our analysis was that it revealed that differences in math SAT or ACT scores and CI pre-scores, which we use as admittedly crude proxies of incoming preparation, were sufficient to explain the performance gaps between demographic groups in our data. Thus, it would be misleading and potentially harmful to discuss gaps in performance between females and males, URM and majority students, and first-generation and continuing generation students, as is customarily done, when the differences in performance are not directly arising from causes associated with those distinctions, but rather appear to be due to differences in incoming preparation. The distinctions in performance are between students with good preparation in physics and poor preparation in physics, or more specifically, good math SAT or ACT scores and CI pre-scores, and poor scores on those two measures. This distinction is the same across all demographic groups.

Addressing a range of incoming preparation is a challenge faced by every physics instructor. How can an instructor best address the range of students in their class in their instruction? Since no institution could (or should!) base its admissions entirely on physics preparation, there is an inherent range of physics preparation in every class. Also, when a student does poorly, how much of that is due to weaknesses in their preparation relative to other students, how much is the result of instruction, and how much is due to other possible factors, such as student demographics and the relationship to social-psychological issues that an instructor may or may not be able to affect?

This paper is following the increasing, but still relatively new, trend to explore questions about factors correlated with student performance using more extensive statistical analyses such as multiple regressions and structural equation modeling (SEM) [21-28]. In an introductory physics course many factors can contribute to student performance, and some of these factors may not act independently—they may interact in complex ways. This can only be explored using multivariable regression analysis. Furthermore, SEM can provide additional information by testing for potential structural relations, such as mediation pathways, between multiple different factors.

The research questions to be explored in this paper are largely empirical:

1. How much of the variation in performance in the standard introductory calculus-based college physics course (physics 1) at three institutions can be explained by readily obtained measures of incoming student characteristics?
2. What are some underlying mechanisms for gender, FG, and URM academic performance gaps in physics 1 and do those justify the singling out of these particular gaps?
3. How similar are the answers to (1) and (2) across different institutions? (The limited sample available for this work only allows a start at answering this question.)

We are not providing answers to these questions that apply to all populations of college students, but rather preliminary observations that we hope will stimulate others to carry out similar analyses so that a larger body of data spanning more institutions and student populations can be accumulated to provide more generalizable answers. Such data is necessary to address the more fundamental question that we and many others find particularly important, namely, what forms of instruction are the most effective at achieving success for the maximum number of students in our courses, given the inherent differences present in any student population?

## Methods

We looked at the large physics 1 course at three large, research-intensive institutions: a highly selective east coast university (HSEC), a highly selective west coast university (HSWC) and a large public research university in the middle of the country (PM). Physics 1 is the standard introductory course that is offered by most physics departments and is taken primarily by students intending to major in engineering, as well as some chemistry and physics majors and some pre-medical students. In Table I we list some characteristics of the institutions and the students in physics 1 at the three institutions.

The data we have for all three institutions are: gender, URM and FG status, proxies for students' incoming preparation (their pre- and post- physics concept inventory test scores, a mix of math SAT or ACT scores), and their course performance (physics 1 final exam scores). Both HSWC and PM used the Force and Motion Conceptual Evaluation (FMCE), while HSEC used the Force Concept Inventory (FCI) as a physics concept inventory [29-30]. Students were considered URM if they were non-white, non-Asians, and were considered first-generation college attending if neither of their parents had a four-year college degree. As we had a mixture of ACT and SAT scores, we converted all of them to percentile scores using available conversion tables, and used the resulting percentile scores in our regression models [31].

In addition, we have particular pieces of data for only one or two institutions. We included these in our model analyses for those institutions to test for the importance. These data include: taking a supplementary help session targeting students with weaker preparation (HSWC, 2018), number that had taken AP physics (HSWC and HSEC), and midterm exam scores (HSWC, 2017). Although we have the physics 1 course grades for all the institutions, we only used final exam scores in our analysis because the grading standards and the course components that go into the calculation of these course grades varied greatly across the three institutions. The structure, administration, and grading of the final exams were similar.

| Institutional characteristics | HSEC | HSWC | PM |
|---|---|---|---|
| # students/yr. taking physics 1 | 194 (2012) | 466 (2017) | 4 offerings 2015-17 |
| | 185 (2013) | 518 (2018) | ~ 1100 per class |
| Math SAT top 25th & 75th % | 790, 700 | 800, 730 | 690, 570 |
| % in top 10% of HS class | 86 | 96 | 29 |
| Physics 1 class characteristics | | | |
| Average percentile math SAT/ACT | 97 | 97 | 89 |
| Average pre-score on concept inventory (%) | 63, 61 | 58, 53 | 38 to 49 |
| Normalized pre-post gain on CI | 0.40, 0.36 | 0.44, 0.47 | 0.49 to 0.54 |
| Institutional characteristics | HSEC | HSWC | PM |

Table I. Institutional characteristics and characteristics of students in physics 1

We carried out multivariable linear regression analyses for each of the three institutions in the data set, using gender, FG, and URM status, as well as various measures of incoming preparation to predict the final exam score. We normalized all the continuous variables in these analyses in terms of the sample standard deviation ("z-scores"), so the coefficients in the models can be directly interpreted as the fraction of a standard deviation in the outcome variable for a one standard deviation change in the continuous predictive variable. For the categorical variables, such as gender, the coefficient refers to the effect of changing from 0 to 1. In these analyses, we examined which combination of the aforementioned variables would provide the simplest, best fitting model to predict students' final exam scores.

In Appendix A, we provide more details as to how the models are evaluated and the criteria used to include terms to find the simplest, best fitting model. In the model evaluation, we focused on the value of R-squared = Explained variation of the outcome variable/ Total variation of the outcome variable (the larger the R-squared, the better the model fit), and the value of the Akaike Information Criterion (AIC) [32]. The AIC is a standard criterion for evaluating the quality of a predictive model that takes into account the parsimony of the model, so considers the number of variables in the model as well as its predictive power (the smaller the AIC score, the better the model).

With regression analysis, one can explore which and to what extent predictors correlate with the dependent variable of interest, but cannot directly explore the relationship between the predictors themselves. To determine these relationships, and the corresponding effects on the dependent variable, student exam performance, we employed structural equation modeling (see Appendix B) using the Lavaan package in R [33-34]. In the structural equation modeling, we tested whether incoming preparation is a mediator for the effect of demographic status on student performance. In other words, we tested whether students from different demographic status have different levels of incoming preparation, and how these differences lead to differences in exam scores.

## Results

**Regression Analysis:**

If one looks only at average exam scores for the various demographic groups, there are significant differences as shown in Table II. This analysis gives an estimate of demographic gaps without controlling for incoming preparation of students from different demographic groups. However, when using multivariable regression to control for student incoming preparation as measured by CI pre-scores and math SAT/ACT scores, the direct effects of demographic variables on student outcomes become insignificant. The coefficients of demographic status in this regression analysis give an estimate of demographic gaps when controlling for incoming preparation as measured by math SAT/ACT and CI pre-score. We illustrate this explicitly in Fig. 1, as well as Table II. The blue (left-most) columns in Fig. 1 show the coefficient of the demographic status for a model that predicts final exam scores including only the single demographic variable (equivalent to a simple t-test); these correspond to the top rows of Table IIa-IIc. The teal (central) columns then show the size of this coefficient in the model when math SAT/ACT scores are added to the model (second rows of Tables IIa-c), and finally, the yellow (right-most) columns show the size of the demographic coefficient after the CI pre-scores (indicating subject specific preparation) are also added to the model as well as math SAT/ACT (third row of the tables).

| a. Predictor of final exam score | HSEC | HSWC | PM |
|---|---|---|---|
| Gender | $b_{Gender} = -0.24\ (0.10)^{**}$<br>$R^2 = 0.01$ | $b_{Gender} = -0.26\ (0.07)^{***}$<br>$R^2 = 0.02$ | $b_{Gender} = -0.28\ (0.04)^{***}$<br>$R^2 = 0.02$ |
| Math SAT/ACT + Gender | $b_{Math} = 0.28\ (0.05)^{***}$<br>$b_{Gender} = -0.26\ (0.10)^{***}$<br>$R^2 = 0.09$ | $b_{Math} = 0.18\ (0.03)^{***}$<br>$b_{Gender} = -0.24\ (0.07)^{***}$<br>$R^2 = 0.08$ | $b_{Math} = 0.37\ (0.02)^{***}$<br>$b_{Gender} = -0.23(0.04)^{***}$<br>$R^2 = 0.20$ |
| Math SAT/ACT + CI + Gender | $b_{CI} = 0.34\ (0.05)^{***}$<br>$b_{Math} = 0.2\ (0.07)^{***}$<br>$b_{Gender} = -0.04\ (0.10)$<br>$R^2 = 0.18$ | $b_{CI} = 0.44\ (0.03)^{***}$<br>$b_{Math} = 0.18\ (0.03)^{***}$<br>$b_{Gender} = -0.018\ (0.06)$<br>$R^2 = 0.30$ | $b_{CI} = 0.38\ (0.02)^{***}$<br>$b_{Math} = 0.25\ (0.02)^{***}$<br>$b_{Gender} = -0.02\ (0.04)$<br>$R^2 = 0.27$ |

| b. Predictor of final exam score | HSEC<br>11-13 | HSWC<br>17-18 | PM<br>14-17 |
|---|---|---|---|
| URM | $b_{URM} = -0.51(0.13)^{***}$<br>$R^2 = 0.03$ | $b_{URM} = -0.38\ (0.11)^{***}$<br>$R^2 = 0.03$ | $b_{URM} = -0.16\ (0.05)^{***}$<br>$R^2 = 0.004$ |
| Math SAT/ACT + URM | $b_{Math} = 0.24\ (0.06)^{***}$<br>$b_{URM} = -0.18\ (0.16)$<br>$R^2 = 0.07$ | $b_{Math} = 0.46\ (0.05)^{***}$<br>$b_{URM} = -0.053\ (0.10)$<br>$R^2 = 0.22$ | $b_{Math} = 0.38\ (0.02)^{***}$<br>$b_{URM} = -0.002\ (0.04)$<br>$R^2 = 0.15$ |
| Math SAT/ACT + CI + URM | $b_{CI} = 0.34\ (0.05)^{***}$<br>$b_{Math} = 0.16\ (0.06)^{**}$<br>$b_{URM} = -0.16\ (0.15)$<br>$R^2 = 0.18$ | $b_{CI} = 0.37\ (0.05)^{***}$<br>$b_{Math} = 0.31\ (0.05)^{***}$<br>$b_{URM} = -0.020\ (0.10)$<br>$R^2 = 0.33$ | $b_{CI} = 0.38\ (0.02)^{***}$<br>$b_{Math} = 0.26\ (0.02)^{***}$<br>$b_{URM} = -0.02\ (0.04)$<br>$R^2 = 0.28$ |

| c. Predictor of final exam score | HSEC | HSWC | PM |
|---|---|---|---|
| FG | $b_{FG}$ = -0.24(22) <br> $R^2$ = 0.0005 | $b_{FG}$ = -0.53 (0.13)*** <br> $R^2$ = 0.04 | $b_{FG}$ = -0.38 (0.05)*** <br> $R^2$ = 0.02 |
| Math SAT/ACT + FG | $b_{Math}$ = 0.27 (0.05)*** <br> $b_{FG}$ = -0.15 (0.22) <br> $R^2$ = 0.07 | $b_{Math}$ = 0.45 (0.05) *** <br> $b_{FG}$ = -0.16 (0.12) <br> $R^2$ = 0.22 | $b_{Math}$ = 0.38(0.02) *** <br> $b_{FG}$ = -0.17 (0.05) *** <br> $R^2$ = 0.15 |
| Math SAT/ACT + CI + FG | $b_{CI}$ = 0.34 (0.05)*** <br> $b_{Math}$ = 0.19 (0.05)*** <br> $b_{FG}$ = -0.12 (0.21) <br> $R^2$ = 0.18 | $b_{CI}$ = 0.37 (0.05)*** <br> $b_{Math}$ = 0.30 (0.05) *** <br> $b_{FG}$ = -0.12 (0.11) <br> $R^2$ = 0.33 | $b_{CI}$= 0.38 (0.02)*** <br> $b_{Math}$ = 0.25 (0.02) *** <br> $b_{FG}$ = -0.11(0.04)* <br> $R^2$ = 0.28 |

**Table II.** Various regression models comparing the effects of incoming preparation and (a) Gender, (b) URM status, or (c) FG status on final exam across the different institutions. The regression coefficients are normalized such that they may be interpreted as an effect size. *** $p < 0.001$, ** $p < 0.01$, * $p < 0.05$ with no correction to p-values for multiple comparisons. As this represents nine regression analyses performed on each of the three data sets, correcting the p-values for multiple comparisons makes the coefficient associated with PM FG status statistically consistent with zero. The p-values for $b_{Math}$ and $b_{CI}$ are so small that the significance of these coefficients is not affected by correction for multiple comparisons. All values of $R^2$ shown are adjusted $R^2$ values.

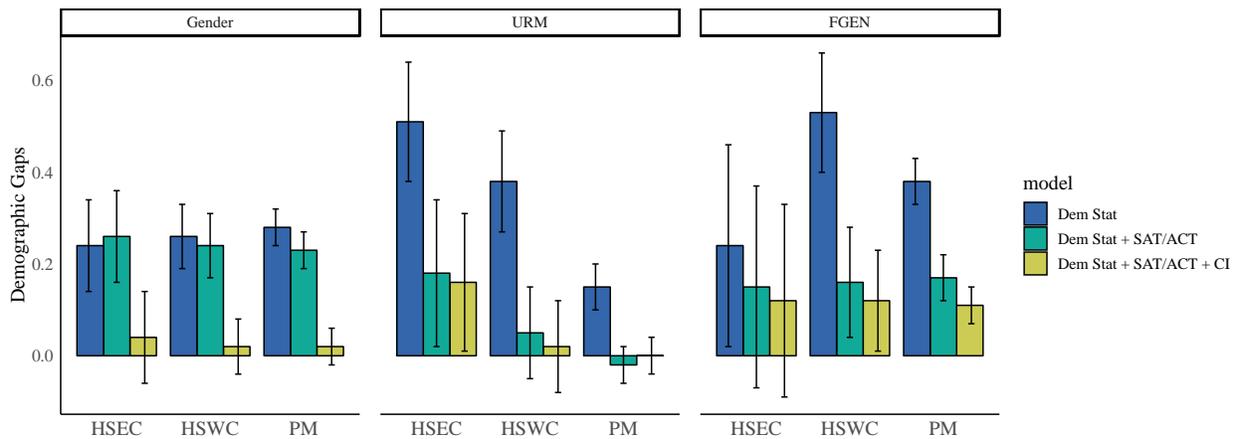

Figure 1: Size of coefficient for the demographic status as predicted by regression models for each institution. First model (blue) shows the coefficient where only the respective demographic status (gender, URM, FG) is included in the model, second model (teal) shows the demographic coefficient when math SAT/ACT score is added as a predictor, and the third model (yellow) is the coefficient when the CI pre-score is added as a predictor as well as math SAT/ACT. The error bars represent the standard error of the coefficients. As Shown in Table II, the regression models with only demographic status have R-squared values of 0.03 or less, but these increase to 0.2-0.3 when measures of incoming preparation are added to the model.

As the gender (left-most) panel of Fig.1 shows, the gender gap, i.e. the gender coefficient, changes little when math SAT is added to the regression model (the change from blue to teal bar in gender panel), reflecting the fact that there is very little difference between average male and female math SAT/ACT scores. However, there is a large change in the gender coefficient when CI pre-score is added to the regression model (the change from teal to yellow bar in gender panel). This change implies a significant average difference on the CI pre-scores between males and females. For the URM gap, a different pattern is apparent. The size of the URM coefficient in the regression model shows more initial variation across institutions, but for all three institutions, when math SAT/ACT is added to the model, the URM gap is drastically reduced and becomes insignificant. For the FG gap, there is less obvious consistency across institutions, except that adding math SAT/ACT to the model changes the FG gap significantly. Overall, the majority of URM and FG gaps could be explained by math SAT/ACT, while math SAT/ACT was negligible in explaining the gender gap. However, all of the gender gap could be explained by CI pre-score. This observation that these two incoming preparation measures had different explanatory power for different demographic gaps demonstrates the importance of having multiple measures to adequately characterize students' incoming preparation, particularly including subject specific measures. Table III shows the regression models with only incoming preparation variables. As shown, the CI and math SAT/ACT can predict almost one third of the overall variance in the final exam scores. Furthermore, adding demographic variables makes a negligible improvement to the adjusted R-squared of the model.

| Predictor of final exam score | HSEC 11-12, 12-13 | HSWC 17-18 | PM14-17 |
|---|---|---|---|
| Math SAT/ACT percentile + pre CI | $b_{CI}$=0.34(0.05)\*\*\* <br> $b_{Math}$= 0.20 (0.05)\*\*\* <br> $R^2$ = 0.18 | $b_{CI}$= .34(0.05)\*\*\* <br> $b_{Math}$ = 0.35(0.05) \*\*\* <br> $R^2$ = 0.34 | $b_{CI}$= 0.37(0.02)\*\*\* <br> $b_{Math}$= 0.39 (0.02) \*\*\* <br> $R^2$ = 0.27 |

Table III: Models predicting the final exam scores only by the two measures of incoming preparation across the three different institutions. The regression coefficients are normalized such that they may be interpreted as an effect size.

In summary, while demographic performance gaps on the physics 1 final exam exist at these institutions, the gaps can be explained by differences in students' incoming preparation as estimated by the two measures of math SAT/ACT and CI pre-score. These two measures are actually rather crude measures for incoming preparation in physics; therefore, it is striking that they are sufficient to eliminate the significance of the demographic variables.

In Appendix B, we use SEM to examine in more detail how CI pre-scores and math SAT/ACT scores mediate the effect of demographic characteristics on final exam scores. This quantifies how the different demographic gaps are mediated by math SAT/ACT score and CI pre-score. The SEM confirms that demographic gaps in final exam scores are mediated by incoming preparation. This is the same qualitative information presented in Table II and Fig. 1 but provides additional quantitative statistical tests [35].

We have also looked at the possible contributions of several other variables, many of which we had for only a subset of institutions. The composite SAT/ACT score had less predictive power than math SAT/ACT: the AIC of the model with only math SAT/ACT as predictor was lower than the model with only composite SAT/ACT as a predictor (820 as opposed to 837). Also, addition of the composite SAT/ACT to the model with math SAT/ACT and CI pre-score as predictors did not change the $R^2$ of the model (0.30 for both models). For HSWC (2018), we were able to examine the effect of cumulative student university GPA at the start of the course. This effect is significant when adding GPA to the model in Table III—$R^2$ goes from 0.34 to 0.47—an expected result as this captures other factors of students' adjustment to the college academic environment. We also examined the effect of a supplementary weekly help session (HSWC 2018) and found no significant impact on final exam score. The HSWC (2017) student scores on a set of CLASS questions reporting on their self-efficacy, both pre- and post-course, were also found to be negligible predictors. In previous work [20], we had already seen that these were the only items on the CLASS survey that showed significant variation across these populations, but here we see that variation is not correlated with exam performance. We also considered the interaction between demographic variables and these factors that we expected might be significant—for example, if the supplementary weekly help session disproportionately helped URM students. No such two-way interactions were found to be significant, and it was assumed therefore that no higher-order interactions would contribute significantly. For HSEC and PM, we had several years of equivalent data, and we found that the inclusion of a random effect of "year" in the model was not significant—these findings are consistent over time. For both HSEC and PM, there was a standard body of exam questions and a process for ensuring equivalence between the different years, which may have contributed to this consistency.

We also used regression models to predict the CI post course scores, as performance gaps on such tests has been a topic of interest [13]. For all three institutions, the distribution of CI post scores is highly distorted, showing a strong ceiling effect. With such distorted distributions, it is questionable as to how valid regression models will be. Similar to what was reported by Day et al., we found that different types of statistical analyses suggest different conclusions [36-37]. A linear regression model indicates that some demographic variables are statistically significant while others are not, but if we model the natural log transform of CI post-score to address the ceiling effect, the model indicates that none of the demographic variables are significant. We interpret this to mean that the CI post distribution is sufficiently distorted that one cannot obtain statistically reliable results from such analyses. For that reason, we present no analysis and make no claims concerning the CI post-scores. We do note that adding the CI post-score as a predictor in our linear regression model of final exam improves the value of R squared by about 0.1 for all three institutions. This does imply that there is considerable overlap in what the CI and the final exams are measuring, even though they appear to have little resemblance.

**Failure analysis:**

We also carried out a "failure analysis," for the students at HSWC in 2018. Results will be similar for the other institutions to the extent the coefficients in Table III are similar. We used the coefficients of the simplest best-fitting model reported in Table III to predict students' final exam score, and then divided these predicted scores into two groups: bottom quartile ("failing"), and top three quartiles ("passing"). The bottom quartile approximately represents a grade cutoff below which students would no longer be meeting the requirements for some programs and majors. We compare the probability of being in the bottom quartile of exam scores for the top and bottom quartiles of preparation. This illustrates the implications of preparation differences where they are the most educationally important: a failing grade often directly impacts a student's future choice of major, and hence likely their career, as well as their probability of graduating.

To quantify incoming preparation, we compute the sum of normalized SAT/ACT and CI pre-scores, weighted by the regression coefficients given in Table III. The results of this analysis are shown in Fig 2. Even though the model only explains 34% of the variance, this failure analysis shows that a student who comes in with preparation in the bottom quartile has about a factor of five higher probability of being in the bottom quartile of the grade distribution than a student who starts the course in the upper quartile of preparation. The distribution of actual student scores between these quartiles is similar to this model prediction, showing that the model is accurately characterizing the actual distribution.

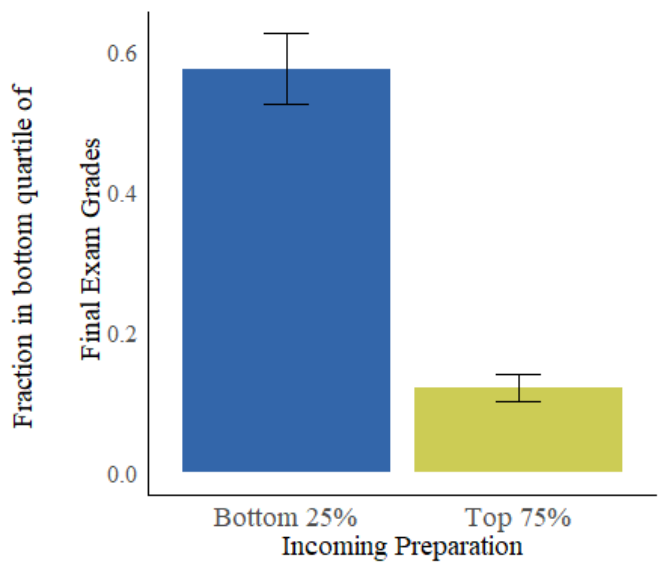

Figure 2: Probability of a student scoring in the bottom 25% of the class as a function of their preparation as measured by the weighted sum of their math SAT/ACT scores and CI pre-scores. Error bars indicate the standard error of the measurement.

## Discussion

The first notable result of this analysis is the degree of similarity across three different institutions in spite of having different admissions criteria and selectivity and locally defined physics courses and final exams. Comparing HSWC and PM, we see substantial differences in both average math SAT/ACT and CI pre-scores, but they have very nearly identical predictive power. The variation in final exam scores predicted by these two factors is about 30% in both cases. Although HSEC is using a different CI, the FCI rather than the FMCE, the final model looks fairly similar to the others, with the same coefficient for the CI term and a 30% smaller coefficient for math SAT/ACT, and a somewhat smaller value of R-squared (0.2 rather than 0.3).

It is natural to wonder, particularly given the similarity of the best predictive models across institutions, as to how similar are the exams and the teaching methods in use at these different institutions. A detailed analysis is beyond the scope of this paper, but we can provide some general observations. Looking at the final exams, they appear rather similar, with the PM exam being slightly easier in terms of complex quantitative calculations and having somewhat more emphasis on basic concepts. In terms of teaching methods, the HSEC course was largely traditional in all aspects (teaching methods have since been modified), while PM was quite interactive. Peer Instruction was used extensively in lectures, and recitation sections used Tutorials in Introductory Physics [38] or similar active learning approaches.

HSWC is in between, with similar activities in section to PM, and limited use of peer instruction in lectures.

The second notable, and arguably most important, result in this paper, is what it says about the gaps in performance associated with demographic characteristics. It shows that it is misleading to do simple t-test comparisons of different demographic groups such as male-female, or URM and majority students. To properly understand the variations in student performance across different demographic groups, it is necessary to do regression analyses taking into account incoming preparation, and those measures of incoming preparation need to include both general levels of preparation and subject specific measures. Such a regression analysis provides an entirely different picture from the t-test. For example, across all three institutions, the initial size of the gender gap as predicted by a single variable model was very similar, about 0.2 standard deviations. When we controlled for students' general incoming preparation as measured by math SAT/ACT score, there was little change in the gap. However, when we also controlled for a subject specific measure of incoming measure, CI pre-score, the gender gap became insignificant for all three institutions. This implies that for these three institutions, once one takes into account differences in students' physics-specific incoming preparation, there is no statistically significant gender gap. For URM gaps, the size of the gap as predicted by single variable model varied substantially across institutions. However, when we controlled for students' math SAT/ACT, the gap became insignificant for all three universities. For FG gaps, the respective sizes of the gaps were quite different at each institution, but controlling for math SAT/ACT scores nearly eliminated the gaps and the differences between institutions.

To emphasize, there are small—if any—gaps in performance associated with demographic differences. There are only performance differences associated with differences in incoming preparation as measured by two proxies: math SAT/ACT and CI pre-score. It is notable that math SAT/ACT and CI scores are two rather crude proxies for incoming preparation. The math SAT covers a variety of math knowledge, but little if any of that seems to be an important component to students' success in physics 1 for many of these students. We have explored students use of math in physics 1 at HSWC in particular, and there was no indication that their performance was limited by math skills. They have mastered all the math they need in physics 1, although their application of math to physical situations is often weak. That is not tested in the math SAT however, so we attribute the significance of the math SAT/ACT score not to math per se, but rather how this score represents some broader level of math-science preparation.

The FCI and FMCE probe a very limited aspect of physics mastery that is needed in physics 1. They test mastery of a limited set of the physics concepts covered, and they are entirely nonmathematical, so they probe nothing about quantitative reasoning and calculational skills which are used extensively on the final exams. Nevertheless, this work shows that it is important to have both general and physics specific measures of incoming preparation, and that rather crude proxies for each of these is sufficient to explain the apparent demographic differences in performance. We cannot identify what factors are important in determining the level of incoming preparation. We initially expected that it would be differences in what high school physics courses were taken, but we analyzed that for HSWC, and we found that all demographic groups at this institution had the same distribution of taking AP physics, regular high school physics, and no physics, even though the groups had different average CI pre-scores and math SAT/ACT scores.

Other analyses that looked at gender differences in physics courses without considering incoming preparation have been used to argue for the importance of social-psychological effects, such as stereotype threat, on the performance of women students [13]. Our analysis, however, implies that such factors are likely not having a significant differential impact on women performance, at least for the student populations we have considered. This is consistent with the findings of Kost et al. [25-26]. This conclusion is also supported by the lack of a correlation between final exam scores and our attitudinal measures of self-efficacy.

## Conclusion

We have examined the variations in the final exam scores in physics 1 across three institutions. This course is a pre-requisite for many engineering and science fields, and therefore demographic performance gaps in the course could be consequential in perpetuating the underrepresentation of some demographic groups in STEM fields. We observed significant demographic gaps in final exam scores for all three institutions. However, when we controlled for students' incoming preparation, in all cases the gaps became insignificant or drastically reduced in size. We find that only incoming math SAT/ACT scores and concept inventory pre-scores together predict 20-30% of the variation in final exam scores. This is surprisingly consistent across three rather different institutions. Similar analysis from a broader range of institutions is needed to determine the generality of these observations. This will allow further studies on the extent to which different teaching methods might reduce the effects of differences in preparation.

The fraction of the variance explained by the two measures of preparation we have used is substantial, but much less than 1, indicating that there are other important variables in student success. Some students with apparently weak preparation still do quite well. We are carrying out further studies to find out what are these important "hidden variables" that determine the rest of the variance. It should also be noted that the analysis in this paper is correlational. While it is plausible that weak preparation causes low exam performance, this does not demonstrate that. It is possible that there is some as yet unknown factor that causes both lower scores on our measures of incoming preparation and lower final exam performance.

We hope that the analysis presented here will stimulate others to collect and publish similar results to provide a baseline to better understand the factors contributing to student performance. Understanding these factors can further help us design instructional practices that will benefit more students, including underrepresented minorities. This work shows that incoming preparation is a major predictor for student performance in Physics1 course, and when controlling for incoming preparation, there remain no demographic performance gaps. Therefore, if we want to improve the outcomes of students from different demographic groups, we have to better address the variation in incoming preparation for all students. This work shows that creating instruction that enhances success of every student across the full range of incoming preparations is also the solution to eliminating gaps in the performance across demographic groups.

Future work will determine how to best do this, but we can offer some potential suggestions. Better matching the introductory course to the range of background preparations of the student population would likely ensure that many more students, particularly those who had the misfortune to attend K-12 schools that provided weaker education in STEM in general and physics in particular, would achieve better outcomes.

In our brief examination of the physics 1 course at the three institutions, it appears that the level and pace of the course is primarily targeted towards the better-prepared students in the distribution, making the course particularly challenging for students with less preparation, and hence, their results are more sensitive to their preparation level. It is plausible that adjusting the course level to better match the preparation of the less prepared students would improve their performance and reduce the sensitivity to preparation, while having a very small impact on the learning of the best prepared. Another option would be to provide greater resources in the teaching of the course, such as classes with more instructor time, or adding courses to the sequence to provide a greater range of students the opportunity to start with a course matched to their preparation. Of course, that would require additional resources, but in these institutions and a number of others we have examined, the amount of resources expended per credit hour in the science disciplines is far greater for the upper level students than for lower level students such as considered here. If it is an institutional priority to maximize the diversity of a student body that is successfully pursuing STEM careers, reversing that inequality in the expenditure of educational resources between upper and lower level courses would very likely help.

## Acknowledgements.

The authors would like to acknowledge the students and instructors from the courses studied, as well as institutional research helpers for providing access to data.

## Appendix A

For model evaluation, one should look for the simplest best-fitting model: the model that has the best fit with the least number of variables (parsimony). One should add more variables to a regression model only if that addition would improve the model fit significantly, i.e. if that addition would significantly increase the percentage of variance of dependent variable that can be explained by the model ($R^2$ of the model). One uses ANOVA to statistically compare the fit of multiple nested models. Models are nested if variables included in a simpler model are a subset of variables included in the more complex model(s). If the models are not nested, then one can compare the values of the AIC index of the models. The smaller the AIC of a model, the better the model fit is. Variables can also produce a statistically significant improvement of the model fit without having practical educational significance. For example, an additional variable that changes the R-squared of the model from 0.29 to 0.30 has little practical significance.

The following tables capture this model evaluation process. Tables AIa-c show different regression models for predicting final exam score using gender, math SAT/ACT, pre-test CI, and different interaction terms between these main factors. Each column represents a model. Each filled cell in a column represents a coefficient of an included variable in the model, with standard deviation of the coefficient presented in parenthesis. Below each column, AIC and adjusted $R^2$ of the model along with other model statistics are reported. Tables were created using the Stargazer package in R [39]. First, we started with single-variable regression models to predict the final exam score, e.g. predicting final exam only by gender. Then we used two-variable regression models to predict the final exam score. In the next step, we use all the three basic factors of gender, math SAT/ACT, and pre-test CI to predict final exam. Finally, to this basic additive model, we added two-way interactions as well as the three-way interactions between all the factors, and tested whether any of these additions improved the model fit. The only interaction term found to improve the model significantly was an interaction between CI pre-score and math SAT/ACT score at PM, but there is no educational significance to this finding.

Results

*Dependent variable:*

Final Exam Score

| | Model1 (1) | Model2 (2) | Model3 (3) | Model4 (4) | Model5 (5) | Model6 (6) | Model7 (7) | Model8 (8) | Model9 (9) | Model10 (10) | Model11 (11) | Model12 (12) | Model13 (13) | Model14 (14) |
|---|---|---|---|---|---|---|---|---|---|---|---|---|---|---|
| Gender (female=1, male=0) | -0.242** (0.103) | | | -0.261*** (0.099) | 0.006 (0.101) | | -0.038 (0.100) | -0.038 (0.100) | -0.036 (0.100) | -0.034 (0.100) | -0.036 (0.100) | -0.034 (0.100) | -0.032 (0.100) | -0.024 (0.103) |
| Math SAT/ACT | | 0.275*** (0.050) | | 0.279*** (0.050) | | 0.196*** (0.048) | 0.198*** (0.049) | 0.196*** (0.065) | 0.199*** (0.049) | 0.229*** (0.054) | 0.194*** (0.066) | 0.198*** (0.065) | 0.231*** (0.054) | 0.203*** (0.066) |
| Pre-test CI | | | 0.385*** (0.048) | | 0.386*** (0.050) | 0.342*** (0.048) | 0.336*** (0.051) | 0.336*** (0.051) | 0.317*** (0.069) | 0.335*** (0.051) | 0.319*** (0.071) | 0.337*** (0.051) | 0.313*** (0.069) | 0.325*** (0.071) |
| Gender * Math SAT/ACT | | | | | | | | 0.019 (0.095) | | | 0.010 (0.098) | 0.090 (0.105) | | 0.066 (0.119) |
| Gender * Pre-test CI | | | | | | | | | 0.039 (0.099) | | 0.036 (0.103) | | 0.046 (0.099) | 0.027 (0.103) |
| Math SAT/ACT * Pre-test CI | | | | | | | | | | 0.059 (0.042) | | 0.076 (0.046) | 0.060 (0.042) | 0.094 (0.070) |
| Gender * Math SAT/ACT * Pre-test CI | | | | | | | | | | | | | | -0.034 (0.094) |
| Intercept | 0.110 (0.069) | -0.002 (0.050) | -0.0004 (0.048) | 0.117* (0.067) | -0.003 (0.066) | -0.002 (0.047) | 0.015 (0.065) | 0.014 (0.065) | 0.020 (0.067) | -0.0003 (0.066) | 0.020 (0.067) | -0.006 (0.066) | 0.006 (0.068) | -0.007 (0.070) |
| AIC | 1072.2 | 1046.9 | 1016.8 | 1042 | 1018.8 | 1000.9 | 1002.7 | 1004.7 | 1004.6 | 1002.7 | 1006.6 | 1004 | 1004.5 | 1007.8 |
| Observations | 378 | 377 | 378 | 377 | 378 | 377 | 377 | 377 | 377 | 377 | 377 | 377 | 377 | 377 |
| $R^2$ | 0.015 | 0.074 | 0.149 | 0.091 | 0.149 | 0.185 | 0.185 | 0.185 | 0.185 | 0.189 | 0.185 | 0.191 | 0.190 | 0.191 |
| Adjusted $R^2$ | 0.012 | 0.071 | 0.147 | 0.086 | 0.144 | 0.180 | 0.178 | 0.176 | 0.177 | 0.181 | 0.174 | 0.180 | 0.179 | 0.176 |
| Residual Std. Error | 0.994 (df = 376) | 0.965 (df = 375) | 0.924 (df = 376) | 0.957 (df = 374) | 0.925 (df = 375) | 0.907 (df = 374) | 0.908 (df = 373) | 0.909 (df = 372) | 0.909 (df = 372) | 0.906 (df = 372) | 0.910 (df = 371) | 0.907 (df = 371) | 0.907 (df = 371) | 0.909 (df = 369) |
| F Statistic | 5.557** (df = 1; 376) | 29.919*** (df = 1; 375) | 65.765*** (df = 1; 376) | 18.666*** (df = 2; 374) | 32.797*** (df = 2; 375) | 42.360*** (df = 2; 374) | 28.223*** (df = 3; 373) | 21.122*** (df = 4; 372) | 21.157*** (df = 4; 372) | 21.717*** (df = 4; 372) | 16.883*** (df = 5; 371) | 17.509*** (df = 5; 371) | 17.381*** (df = 5; 371) | 12.475*** (df = 7; 369) |

*Notes:*                                                                                           *p<0.1; **p<0.05; ***p<0.01

Table AIa. Different regression models fitted to HSEC data to predict students' final exam score.

**Results**

|  | | | | | | Dependent variable: | | | | | | | | |
|---|---|---|---|---|---|---|---|---|---|---|---|---|---|---|
|  | | | | | | FinalExam Score | | | | | | | | |
|  | Model1 (1) | Model2 (2) | Model3 (3) | Model4 (4) | Model5 (5) | Model6 (6) | Model7 (7) | Model8 (8) | Model9 (9) | Model10 (10) | Model11 (11) | Model12 (12) | Model13 (13) | Model14 (14) |
| Gender (female=1, male=0) | -0.257*** (0.071) |  |  | -0.217*** (0.066) | -0.013 (0.065) |  | -0.035 (0.063) | -0.035 (0.063) | -0.031 (0.063) | -0.033 (0.063) | -0.032 (0.063) | -0.033 (0.063) | -0.029 (0.063) | 0.001 (0.071) |
| Math SAT/ACT |  | 0.375*** (0.033) |  | 0.369*** (0.033) |  | 0.220*** (0.033) | 0.221*** (0.033) | 0.186*** (0.043) | 0.220*** (0.033) | 0.186*** (0.040) | 0.197*** (0.044) | 0.160*** (0.047) | 0.184*** (0.040) | 0.184*** (0.050) |
| Pre-test CI |  |  | 0.480*** (0.031) |  | 0.478*** (0.032) | 0.393*** (0.033) | 0.388*** (0.034) | 0.387*** (0.034) | 0.347*** (0.044) | 0.400*** (0.035) | 0.355*** (0.046) | 0.397*** (0.035) | 0.358*** (0.045) | 0.359*** (0.046) |
| Gender * Math SAT/ACT |  |  |  |  |  |  |  | 0.080 (0.061) |  |  | 0.052 (0.067) | 0.067 (0.062) |  | -0.016 (0.086) |
| Gender * Pre-test CI |  |  |  |  |  |  |  |  | 0.095 (0.063) |  | 0.073 (0.069) |  | 0.094 (0.063) | 0.096 (0.072) |
| Math SAT/ACT * Pre-test CI |  |  |  |  |  |  |  |  |  | -0.058 (0.037) |  | -0.051 (0.038) | -0.057 (0.037) | -0.027 (0.047) |
| Gender * Math SAT/ACT * Pre-test CI |  |  |  |  |  |  |  |  |  |  |  |  |  | -0.075 (0.078) |
| Intercept | 0.123** (0.049) | -0.000 (0.033) | -0.000 (0.031) | 0.104** (0.046) | 0.006 (0.044) | -0.000 (0.030) | 0.017 (0.043) | 0.019 (0.043) | 0.027 (0.044) | 0.038 (0.045) | 0.026 (0.044) | 0.038 (0.045) | 0.049 (0.046) | 0.036 (0.047) |
| AIC | 2222.5 | 2116.3 | 2029.7 | 2107.5 | 2031.6 | 1988.6 | 1990.3 | 1990.6 | 1990.1 | 1989.9 | 1991.5 | 1990.7 | 1989.7 | 1992.5 |
| Observations | 786 | 786 | 786 | 786 | 786 | 786 | 786 | 786 | 786 | 786 | 786 | 786 | 786 | 786 |
| $R^2$ | 0.017 | 0.141 | 0.230 | 0.153 | 0.231 | 0.271 | 0.272 | 0.273 | 0.274 | 0.274 | 0.274 | 0.275 | 0.276 | 0.277 |
| Adjusted $R^2$ | 0.015 | 0.140 | 0.229 | 0.150 | 0.229 | 0.270 | 0.269 | 0.270 | 0.270 | 0.270 | 0.270 | 0.270 | 0.271 | 0.271 |
| Residual Std. Error | 0.992 (df = 784) | 0.928 (df = 784) | 0.878 (df = 784) | 0.922 (df = 783) | 0.878 (df = 783) | 0.855 (df = 783) | 0.855 (df = 782) | 0.855 (df = 781) | 0.855 (df = 781) | 0.854 (df = 781) | 0.855 (df = 780) | 0.854 (df = 780) | 0.854 (df = 780) | 0.854 (df = 778) |
| F Statistic | 13.181*** (df = 1; 784) | 128.419*** (df = 1; 784) | 234.799*** (df = 1; 784) | 70.449*** (df = 2; 783) | 117.276*** (df = 2; 783) | 145.886*** (df = 2; 783) | 97.274*** (df = 3; 782) | 73.444*** (df = 4; 781) | 73.624*** (df = 4; 781) | 73.691*** (df = 4; 781) | 58.987*** (df = 5; 780) | 59.195*** (df = 5; 780) | 59.485*** (df = 5; 780) | 42.612*** (df = 7; 778) |

Note: *p<0.1; **p<0.05; ***p<0.01

Table AIb. Different regression models fitted to HSWC data to predict students' final exam score.

**Results**

*Dependent variable:*

Final Exam Score

| | Model1 (1) | Model2 (2) | Model3 (3) | Model4 (4) | Model5 (5) | Model6 (6) | Model7 (7) | Model8 (8) | Model9 (9) | Model10 (10) | Model11 (11) | Model12 (12) | Model13 (13) | Model14 (14) |
|---|---|---|---|---|---|---|---|---|---|---|---|---|---|---|
| Gender (female=1, male=0) | -0.297*** | | | -0.244*** | -0.014 | | -0.024 | -0.015 | 0.005 | -0.024 | 0.003 | -0.014 | 0.006 | 0.006 |
| | (0.043) | | | (0.040) | (0.039) | | (0.038) | (0.038) | (0.040) | (0.038) | (0.040) | (0.038) | (0.040) | (0.043) |
| Math SAT/ACT | | 0.372*** | | 0.365*** | | 0.246*** | 0.247*** | 0.211*** | 0.246*** | 0.260*** | 0.215*** | 0.224*** | 0.261*** | 0.228*** |
| | | (0.018) | | (0.018) | | (0.017) | (0.017) | (0.021) | (0.017) | (0.019) | (0.021) | (0.021) | (0.019) | (0.022) |
| Pre-test CI | | | 0.478*** | | 0.477*** | 0.401*** | 0.398*** | 0.403*** | 0.382*** | 0.389*** | 0.392*** | 0.391*** | 0.371*** | 0.380*** |
| | | | (0.017) | | (0.018) | (0.017) | (0.018) | (0.018) | (0.019) | (0.018) | (0.020) | (0.018) | (0.020) | (0.021) |
| Gender * Math SAT/ACT | | | | | | | | 0.106*** | | | 0.094*** | 0.124*** | | 0.111** |
| | | | | | | | | (0.035) | | | (0.036) | (0.036) | | (0.045) |
| Gender * Pre-test CI | | | | | | | | | 0.089** | | 0.059 | | 0.095** | 0.061 |
| | | | | | | | | | (0.044) | | (0.046) | | (0.044) | (0.047) |
| Math SAT/ACT * Pre-test CI | | | | | | | | | | 0.036* | | 0.050** | 0.039** | 0.051** |
| | | | | | | | | | | (0.019) | | (0.020) | (0.019) | (0.022) |
| Gender * Math SAT/ACT * Pre-test CI | | | | | | | | | | | | | | -0.002 |
| | | | | | | | | | | | | | | (0.049) |
| Intercept | 0.082*** | 0.000 | -0.000 | 0.067*** | 0.004 | 0.000 | 0.007 | 0.007 | 0.009 | -0.005 | 0.009 | -0.008 | -0.003 | -0.007 |
| | (0.022) | (0.018) | (0.017) | (0.021) | (0.020) | (0.016) | (0.019) | (0.019) | (0.019) | (0.020) | (0.019) | (0.020) | (0.020) | (0.021) |
| AIC | 7627.8 | 7273.5 | 6973.5 | 7238.2 | 6975.4 | 6776.5 | 6778.1 | 6771 | 6776 | 6776.6 | 6771.3 | 6766.6 | 6774 | 6768.8 |
| Observations | 2,703 | 2,703 | 2,703 | 2,703 | 2,703 | 2,703 | 2,703 | 2,703 | 2,703 | 2,703 | 2,703 | 2,703 | 2,703 | 2,703 |
| $R^2$ | 0.018 | 0.138 | 0.229 | 0.150 | 0.229 | 0.284 | 0.284 | 0.286 | 0.285 | 0.285 | 0.285 | 0.288 | 0.288 | 0.288 |
| Adjusted $R^2$ | 0.017 | 0.138 | 0.229 | 0.149 | 0.228 | 0.283 | 0.283 | 0.285 | 0.284 | 0.284 | 0.285 | 0.286 | 0.284 | 0.286 |
| Residual Std. Error | 0.991 (df = 2701) | 0.928 (df = 2701) | 0.878 (df = 2701) | 0.922 (df = 2700) | 0.878 (df = 2700) | 0.847 (df = 2700) | 0.847 (df = 2699) | 0.846 (df = 2698) | 0.846 (df = 2698) | 0.846 (df = 2698) | 0.845 (df = 2697) | 0.846 (df = 2697) | 0.846 (df = 2697) | 0.845 (df = 2695) |
| F Statistic | 48.403*** (df = 1; 2701) | 433.465*** (df = 1; 2701) | 801.385*** (df = 1; 2701) | 238.387*** (df = 2; 2700) | 400.625*** (df = 2; 2700) | 534.272*** (df = 2; 2700) | 356.242*** (df = 3; 2699) | 270.255*** (df = 4; 2698) | 268.519*** (df = 4; 2698) | 268.289*** (df = 4; 2698) | 216.597*** (df = 5; 2697) | 217.927*** (df = 5; 2697) | 215.843*** (df = 5; 2697) | 155.002*** (df = 7; 2695) |

*Note:* *p<0.1; **p<0.05; ***p<0.01

Table AIc. Different regression models fitted to PM data to predict students' final exam score.

For the HSWC 2018, HSEC, and PM data, we had URM and FG status in addition to gender. Therefore, we conducted regression analysis including these additional factors as well as gender, math SAT/ACT, and CI pre-score. In this analysis, if a factor did not have a significant main-effect contribution, we did not consider it for an interaction term. The following tables show a selection of models fitted to the data. It is clear from the regression tables above and below that predicting final exam scores using math SAT/ACT and CI pre-scores is far better than using demographics variables alone—the former explains three to seven times more of the exam variance across than the latter model the three institutions.

| | Results | | |
|---|---|---|---|
| | *Dependent variable:* | | |
| | Final Exam Score | | |
| | Model1 | Model2 | Model3 |
| | (1) | (2) | (3) |
| Gender (female=1, male=0) | -0.218*** | | -0.042 |
| | (0.072) | | (0.072) |
| URM | -0.398*** | | -0.121 |
| | (0.095) | | (0.105) |
| FG | -0.142 | | -0.078 |
| | (0.155) | | (0.148) |
| Math SAT/ACT | | 0.196*** | 0.162*** |
| | | (0.048) | (0.057) |
| Pre-test CI | | 0.342*** | 0.330*** |
| | | (0.048) | (0.051) |
| Constant | -0.288** | -0.002 | -0.110 |
| | (0.116) | (0.047) | (0.114) |
| AIC | 1056.4 | 999.9 | 1004.1 |
| Observations | 378 | 377 | 377 |
| $R^2$ | 0.062 | 0.185 | 0.189 |
| Adjusted $R^2$ | 0.055 | 0.180 | 0.178 |
| Residual Std. Error | 0.971 (df = 374) | 0.905 (df = 374) | 0.907 (df = 371) |
| F Statistic | 8.287*** (df = 3; 374) | 42.360*** (df = 2; 374) | 17.243*** (df = 5; 371) |
| *Note:* | | | *p<0.1; **p<0.05; ***p<0.01 |

Table AIIa. Different regression models fitted to HSEC data using additional demographic data.

**Results**

| | Dependent variable: | | |
|---|---|---|---|
| | Final Exam Score | | |
| | Model1 | Model2 | Model3 |
| | (1) | (2) | (3) |
| Gender (female=1, male=0) | $-0.298^{***}$ | | $-0.060$ |
| | (0.098) | | (0.086) |
| URM | $-0.251^{**}$ | | $0.024$ |
| | (0.111) | | (0.097) |
| FG | $-0.453^{***}$ | | $-0.111$ |
| | (0.132) | | (0.116) |
| Math SAT/ACT | | $0.340^{***}$ | $0.331^{***}$ |
| | | (0.047) | (0.050) |
| Pre-test CI | | $0.345^{***}$ | $0.336^{***}$ |
| | | (0.047) | (0.048) |
| Constant | $0.309^{***}$ | $-0.000$ | $0.042$ |
| | (0.077) | (0.041) | (0.068) |
| AIC | 1095.2 | 957.3 | 962 |
| Observations | 394 | 394 | 394 |
| $R^2$ | 0.078 | 0.347 | 0.349 |
| Adjusted $R^2$ | 0.071 | 0.344 | 0.341 |
| Residual Std. Error | 0.964 (df = 390) | 0.810 (df = 391) | 0.812 (df = 388) |
| F Statistic | $10.960^{***}$ (df = 3; 390) | $103.835^{***}$ (df = 2; 391) | $41.607^{***}$ (df = 5; 388) |
| *Note:* | | | $^{*}$p<0.1; $^{**}$p<0.05; $^{***}$p<0.01 |

Table AIIb. Different regression models fitted to HSWC 18 data using additional demographic data.

 route.col.sl4tor2.95.html

| | Model1 (1) | Model2 (2) | Model3 (3) | Model4 (4) | Model5 (5) | Model6 (6) |
|---|---|---|---|---|---|---|
| | **Results** | | | | | |
| | *Dependent variable:* | | | | | |
| | Final Exam Score | | | | | |
| Gender (female=1, male=0) | -0.288*** (0.043) | | -0.020 (0.038) | -0.023 (0.038) | -0.024 (0.038) | -0.025 (0.038) |
| URM | -0.111** (0.046) | | -0.003 (0.041) | 0.0001 (0.041) | -0.0002 (0.041) | 0.001 (0.041) |
| FG | -0.388*** (0.051) | | -0.113** (0.045) | -0.142*** (0.047) | -0.136*** (0.047) | -0.121** (0.048) |
| Math SAT/ACT | | 0.259*** (0.017) | 0.250*** (0.018) | 0.278*** (0.021) | 0.280*** (0.022) | 0.298*** (0.023) |
| Pre-test CI | | 0.384*** (0.017) | 0.379*** (0.018) | 0.374*** (0.018) | 0.369*** (0.019) | 0.353*** (0.021) |
| FG * Math SAT/ACT | | | | -0.089** (0.037) | -0.096** (0.038) | -0.115*** (0.043) |
| FG * Pre-test CI | | | | | 0.036 (0.050) | 0.051 (0.052) |
| Math SAT/ACT * Pre-test CI | | | | | | 0.046* (0.024) |
| FG * Math SAT/ACT * Pre-test CI | | | | | | -0.050 (0.046) |
| Intercept | 0.166*** (0.026) | -0.000 (0.016) | 0.026 (0.023) | 0.024 (0.023) | 0.024 (0.023) | 0.009 (0.023) |
| AIC | 7476.8 | 6712.9 | 6712.1 | 6708.3 | 6709.8 | 6710 |
| Observations | 2,669 | 2,669 | 2,669 | 2,669 | 2,669 | 2,669 |
| $R^2$ | 0.039 | 0.278 | 0.280 | 0.281 | 0.281 | 0.282 |
| Adjusted $R^2$ | 0.038 | 0.277 | 0.278 | 0.280 | 0.279 | 0.280 |
| Residual Std. Error | 0.981 (df = 2665) | 0.850 (df = 2666) | 0.850 (df = 2663) | 0.849 (df = 2662) | 0.849 (df = 2661) | 0.849 (df = 2659) |
| F Statistic | 36.177*** (df = 3; 2665) | 512.596*** (df = 2; 2666) | 206.677*** (df = 5; 2663) | 173.504*** (df = 6; 2662) | 148.764*** (df = 7; 2661) | 116.200*** (df = 9; 2659) |

*Note:* *p<0.1; **p<0.05; ***p<0.01

Table AIIc. Different regression models fitted to PM data using additional demographic data.

# Appendix B

We used structural equation modeling (SEM) to test a mediation model for each institution. These models show how math SAT/ACT and CI pre-score mediate the observed differences in final exam scores across demographic groups, as well as the size of the respective mediating effects for each demographic group. In the following models (Figures B1a-c), we first show how gender predicts both math SAT/ACT and CI pre-score. We also show how those two measures of incoming preparation are correlated. Finally, we show how student final exam score is predicted by both math SAT/ACT and CI pre-score. Therefore, the gender effect on final exam score is mediated through both math SAT/ACT, and CI pre-score. This model fits the data well for all institutions, as all the fit indices are within the acceptable range (root mean square error (RMSEA): acceptable range: 0-0.07; comparative fit index (CFI): acceptable range: above 0.95; standardized root mean square residual (SRMR): acceptable range: 0-0.1), and the estimated co-variances by the models were not significantly different from the actual co-variances in the data, as suggested by insignificant $\chi^2$ statistics of the models (the null hypothesis in this case is that the model is a good fit). It is notable that this model does not include any direct effect of gender on exam score, which suggests after controlling for the effect of math SAT/ACT and CI pre-score on final exam, there is no significant gender difference in final exam performance.

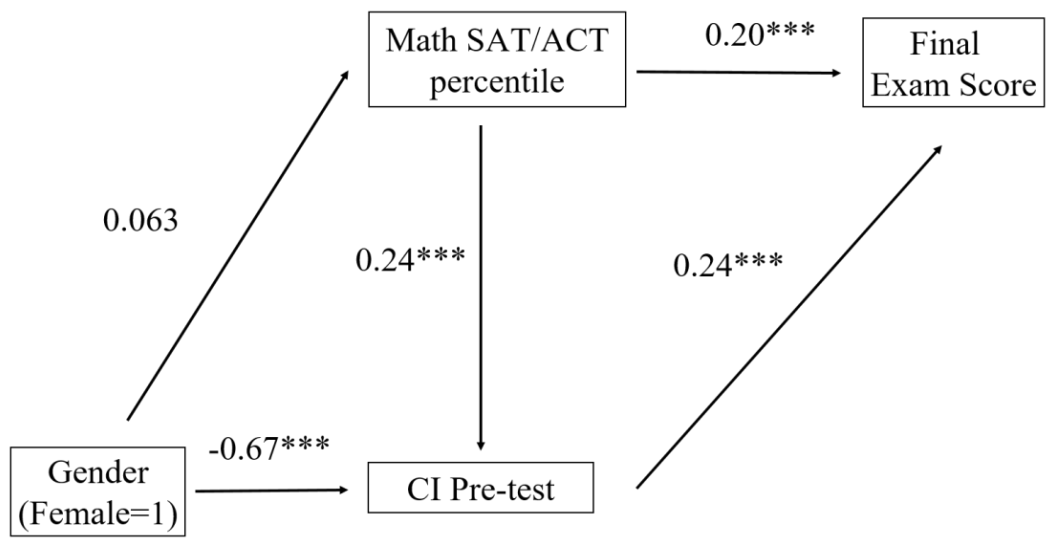

$X2(1)= 0.14$, p = 0.71;  CFI= 1.00; SRMR=0.005; RMSEA= 0.000

Figure B1a- The SEM model for HSEC data.

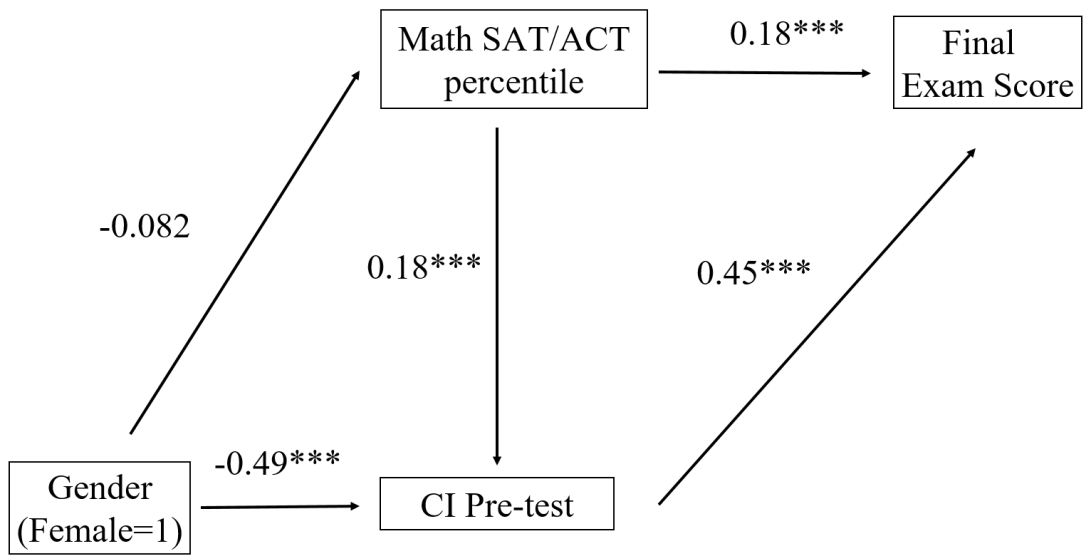

X2(1)= 0.08, p = 0.78;  CFI= 1.00; SRMR=0.003; RMSEA= 0.000

Figure B1b- The SEM model for HSWC data.

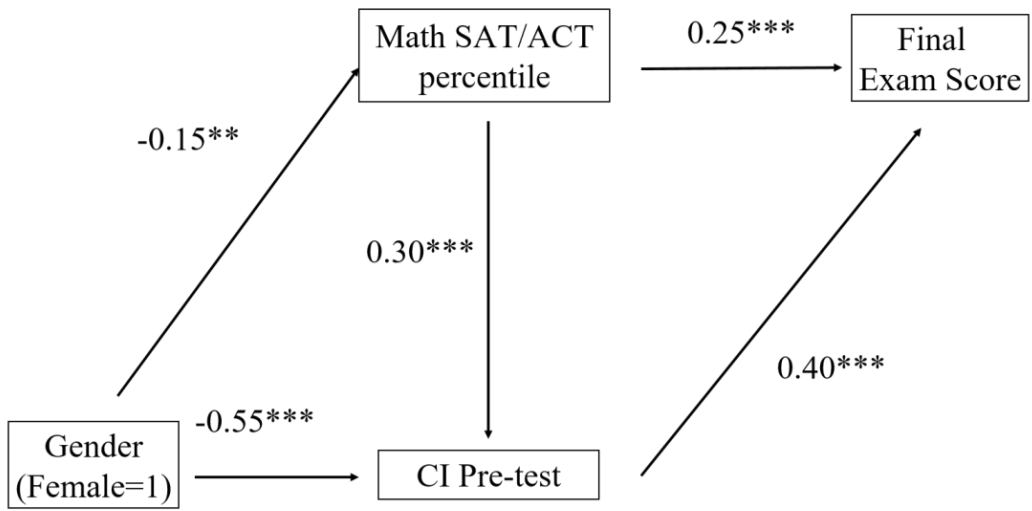

X2(1)=0.42, p = 0.519;  CFI= 1.00. SRMR=0.003, RMSEA= 0.00

Figure B1c- The SEM model for PM data.

We added URM and FG status to the SEM model to test how measures of incoming preparation mediated the effect of these students' demographics on final exam score. Figures B2a-c illustrate SEM analysis including URM, and FG status as well as students' gender. For all the three institutions, the SEM model was a good fit for data, as all the fit indices were within acceptable range, and $\chi^2$ values of the models were insignificant.

Based on the results of these mediation models, for all three institutions, the differences in incoming preparation of under-represented demographic groups (URM, FG, and female students) mediated the difference in their final exam scores. After controlling for the effects of math SAT/ACT and CI pre-score on final exam score, there was no significant difference between the final exam score of under-represented students and their majority peers. One exception was that after controlling for the effects of math SAT/ACT and CI pre-score on exam, there remained a significant but smaller difference between final exam score of FG students and continuing generation students at PM.

For HSEC (Figure S2a), math SAT/ACT and CI pre-score fully mediated the effect of URM on final exam score. URM students had lower math SAT/ACT scores on average, and these lower scores led to lower final exam scores both directly and indirectly via their effect on the CI pre-score. The effect of gender on final exam score was also fully mediated by CI pre-scores. Female students on average had lower CI pre-score. CI pre-score was positively correlated with final exam score, so lower scores on this measure of incoming preparation led to lower final exam scores. For FG students there was no significant gap in performance, and therefore, no mediation effect through measures of incoming preparation.

For HSWC (Figure S2b), math SAT/ACT and CI pre-score fully mediated the effect of URM on final exam score. URM and FG students had on average lower Math SAT/ACT scores, and these lower scores led to lower final exam scores both directly and indirectly via their effect on the CI pre-scores. The effect of gender on final exam score was also fully mediated by incoming preparation, both directly through lower math SAT/ACT scores and CI pre-scores, as well as indirectly by math SAT score via CI pre-score. Female students on average had both lower Math SAT/ACT and lower CI pre-score. Both of these scores were positively correlated with final exam score, so lower scores on these measures of incoming preparation led to lower scores on the final exam. Furthermore, math SAT/ACT score was positively correlated with CI pre-score. Therefore, lower math SAT/ACT scores of female students not only directly mediated the effect of gender on final exam performance, but also indirectly mediated through leading to lower CI pre-score and the lower CI pre-score leading to lower final exam scores.

For PM (Figure B2c), math SAT/ACT and CI pre-score fully mediated the effect of URM and gender on final exam score. URM had on average lower Math SAT/ACT scores, and these lower scores led to lower final exam scores both directly, and indirectly via their effect on the CI pre-score. Both of these scores were positively correlated with final exam scores, so lower scores on these measures of incoming preparation led to lower scores on the final exam. Furthermore, math SAT/ACT score was positively correlated with CI pre-score. Therefore, lower math SAT/ACT scores of female students not only directly mediated the effect of gender on final exam performance, but also indirectly mediated through leading to lower CI-pre scores and these lower CI scores leading to lower final exam scores. The effect of gender on final exam score was also fully mediated by incoming preparation, both directly through lower math SAT/ACT score and CI pre-score, as well as the indirect effect of math SAT score via CI pre-score. Female students on average had both lower Math SAT/ACT and lower CI pre-scores. Both of these scores were positively correlated with final exam scores, so lower scores on these measures of incoming preparation led to lower scores on final exam. Furthermore, math SAT/ACT score was positively correlated with CI pre-score. Therefore, lower math SAT/ACT scores of female students not only directly mediated the effect of gender on final exam performance, but also indirectly mediated through leading to lower CI-pre scores and these lower CI scores leading to lower final exam scores. The effect of FG on final exam was also partially mediated by incoming preparation, both directly through lower math SAT/ACT scores and CI pre-scores, as well as the indirect effect of math SAT scores via CI pre-scores. This was partial mediation, as after controlling for the effect of Math SAT/ACT and CI pre-score on final exam, there existed a significant but smaller FG performance gap in final exam.

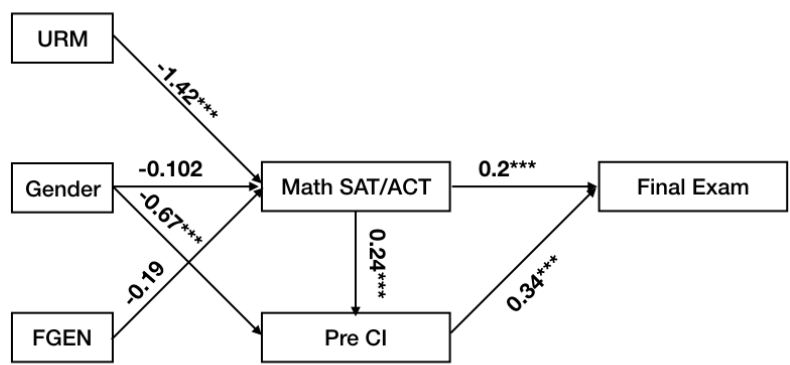

$\chi2(5) = 4.27$, $p = 0.51$; RMSEA = 0.00; SRMR = 0.021; CFI = 1.00

Figure B2a- the SEM model for HSEC data using additional data about students' demographic and incoming preparation

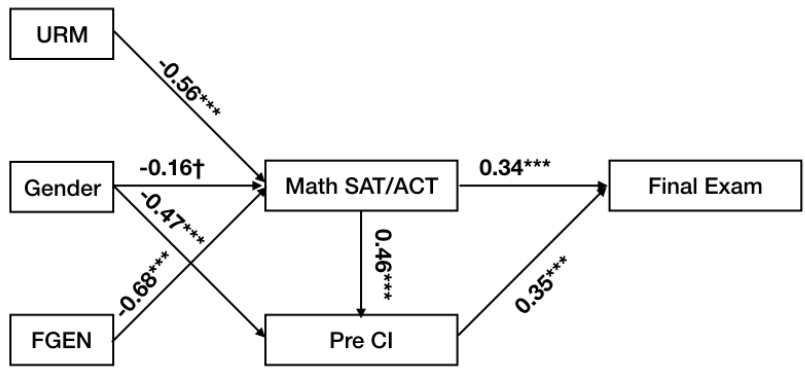

$\chi 2(5) = 1.58$, $p = 0.90$ ; RMSEA = 0.00; SRMR = 0.011; CFI = 1.00

Figure B2b- the SEM model for HSWC 18 data using additional data about students' demographic and incoming preparation

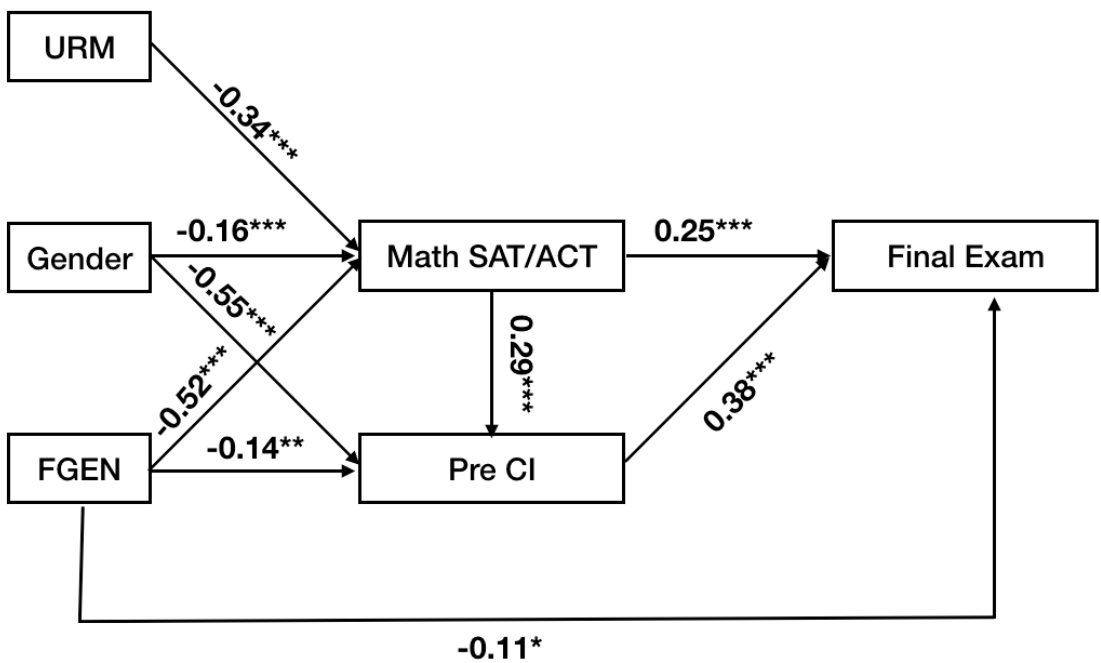

$\chi 2(3) = 0.88$, $p = 0.83$ ; RMSEA = 0.00; SRMR = 0.004; CFI = 1.00

Figure B2c- the SEM model for PM data using additional data about students' demographic and incoming preparation

# References

1. C. H. Crouch & E. Mazur, Peer instruction: Ten years of experience and results, Am. J. Phys. **69**, 970 (2001).
2. K. Cummings, J. Marx, R. Thornton, and D. Kuhl, Evaluating innovation in studio physics, Am. J. Phys. **67**, S38 (1999).
3. R. Hake, Interactive-engagement versus traditional methods: A six-thousand-student survey of mechanics test data for introductory physics courses, Am. J. Phys., **66**, 64 (1998).
4. S. Freeman, S. L. Eddy, M. McDonough, M. K. Smith, N. Okoroafor, H. Jordt, M. P. Wenderoth, Active learning increases student performance in science, engineering, and mathematics, PNAS **111**, 23 (2014).
5. M. Lorenzo, C. H. Crouch, E, Mazur E, Reducing the gender gap in the physics classroom, Am. J. Phys **74**, 118 (2006).
6. R. J. Beichner, J. M. Saul, D. S. Abbott, J. J. Morse, D. Deardorff, R. J. Allain, J. S. Risely, The student-centered activities for large enrollment undergraduate programs (SCALE-UP) project, Research-Based Reform of University Physics 1, 2 (2007).
7. S. Freeman, E. O'Connor, J. W. Parks, M. Cunningham, D. Hurley, D. Haak, M. P. Wenderoth, Prescribed active learning increases performance in introductory biology, CBE Life. Sci. Educ. **6**, 132 (2007).
8. D. C. Haak, J. HilleRisLambers, E. Pitre, S. Freeman, Increased structure and active learning reduce the achievement gap in introductory biology, Science 332, **1213** (2011).
9. C. J. Ballen, C. Wieman, S. Salehi, J. B. Searle, & K. R. Zamudio, Enhancing diversity in undergraduate science: Self-efficacy drives performance gains with active learning, CBE Life Sci. Educ. **16**, 56 (2017).
10. R. Fullilove, & P. Treisman, Mathematics Achievement Among African American Undergraduates at the University of California, Berkeley: An Evaluation of the Mathematics Workshop Program, J. Negro Educ. **59**, 463 (1990).
11. R. H. Tai and P. M. Sadler, Gender differences in introductory undergraduate physics performance: University physics versus college physics in the USA, Int. J. Sci. Educ., **23**, 1017 (2001).
12. T. G. Greene, C. N. Marti, & K. McClennney, The effort-outcome gap: Differences for African American and Hispanic community college students in student engagement and academic achievement, J. Higher Educ. **79**, 513 (2009).
13. A. Madsen, S.B. McKagan, and E. Sayre, Gender gap on concept inventories in physics: What is consistent, what is inconsistent, and what factors influence the gap?, Phys. Rev. Phys. Educ. Res. **9**, 020121 (2013).
14. S. L. Eddy & K. Hogan, Getting under the hood: How and for whom does increasing course structure work?, CBE Life Sci. Educ. **13**, 453 (2014).
15. S. L. Eddy, S. E. Brownell, & M. P. Wenderoth, Gender gaps in achievement and participation in multiple introductory biology classrooms, CBE Life Sci. Educ. **13**, 472 (2014).
16. National Science Foundation; National Center for Science and Engineering Statistics. Women, Minorities, and Persons with Disabilities in Science and Engineering (Special Report) Arlington, VA: National Science Foundation; 2015.
17. S. L. Eddy & S. E. Brownell, Beneath the numbers: A review of gender disparities in undergraduate education across science, technology, engineering, and math disciplines, Phys. Rev. Phys. Educ. Res. **12**, 020106 (2016).
18. J. B. Hinnant, M. O'Brien, & S. R. Ghazarian, The Longitudinal Relations of Teacher Expectations to Achievement in the Early School Years, J. Educ. Psych. **101**, 662 (2009).
19. C. M. Steele & J. Aronson, Contending with a stereotype: African-American intellectual test performance and stereotype threat, J. Pers. and Social Psych. **69,** 797 (1995).
20. L.E. Kost, S.J. Pollock, and N.D. Finkelstein, "Unpacking gender differences in students' perceived experiences in introductory physics," in AIP conference proceedings, Vol. 1179 (AIP, 2009) pp. 177–180.
21. J. Watkins, *Examining issues of underrepresented minority students in introductory physics.* Ph.D. Thesis, Harvard University, 289 pages, (2010).
22. E. Brewe et al., "Toward Equity through Participation in Modeling Instruction in Introductory University Physics", Phys. Rev. Phys. Educ. Res. 6, 0101016 (2010)
23. B. Wilcox and H. Lewandowski, "Research-based assessment of students' beliefs about experimental physics: When is gender a factor", Phs. Rev. Phys. Educ Res. 12, 020130 (2016).
24. R. Henderson, J. Stewart, & A. Traxler, Partitioning the Gender Gap in Physics Conceptual Inventories: FCI, FMCE, and CSEM (2019).

25. L. E. Kost, S. J. Pollock, N. D. & Finkelstein, Characterizing the gender gap in introductory physics, Phs. Rev. ST Phys. Educ. Res. **5**, 010101 (2009).

26. L. E. Kost-Smith, S. J. Pollock, & N. D. Finkelstein, Gender disparities in second-semester college physics: The incremental effects of a "smog of bias," Phys. Rev. ST. Phys. Educ. Res. **6**, 020112 (2010).

27. R. Henderson, G. Stewart, J. Stewart, L. Michaluk, & A. Traxler, Exploring the gender gap in the conceptual survey of electricity and magnetism, Phys. Rev. Phys. Educ. Res. **13**, 020114 (2017).

28. Z. Hazari, R. H. Tai, and P. M. Sadler, Gender differences in introductory university physics performance: The influence of high school physics preparation and affective factors, Sci. Educ. **91**, 847 (2007).

29. R.K. Thornton and D.R. Sokoloff, "Assessing student learning of Newton's laws: The Force and Motion Conceptual Evaluation and the evaluation of active learning laboratory and lecture curricula," Am. J. Phys. 66, 338– 352 (1998).

30. D. Hestenes, M. Wells, and G. Swackhamer, "Force Concept Inventory," Phys. Teach. 30, 141–158 (1992).

31. L. Staffaroni (n.d.). Historical SAT Percentiles: New SAT 2016, 2017, and 2018. Retrieved from https://blog.prepscholar.com/historical-percentiles-new-sat

32. H. Akaike, Likelihood of a model and information criteria, J. Econometrics, 1981 **16**, 3-14. (1981).

33. Y. Rosseel, (2012). Lavaan: An R package for structural equation modeling and more. Version 0.5–12 (BETA). Journal of statistical software, 48(2), 1-36.

34. R Core Team (2013). R: A language and environment for statistical computing. R Foundation for Statistical Computing, Vienna, Austria. URL http://www.R-project.org/.

35. D. Gefen, D. W. Straub, & M. Boudreau, Structural Equation Modeling and Regression: Guidelines for Research Practice, *CAIS,* **4**, 7 (2000).

36. J. Day, J. B. Stang, N. G. Holmes, D. Kumar, & D. A. Bonn, Gender gaps and gendered action in a first-year physics laboratory, Phys. Rev. Phys. Educ. Res. **12**, 020104 (2016).

37. S. D. Willoughby & A. Metz, Exploring gender differences with different gain calculations in astronomy and biology, Am. J. Phys **77,** 651 (2009).

38. McDermott, L. C., & Shaffer, P. S. (2002). *Tutorials in introductory physics*. Upper Saddle River, NJ: Prentice Hall.

39. M. Hlavac, (2018). stargazer: Well-Formatted Regression and Summary Statistics Tables. R package version 5.2.1. https://CRAN.R-project.org/package=stargazer